\documentclass[12pt]{iopart}

\usepackage{amssymb}
\usepackage{graphicx}
\usepackage{bm}
\usepackage{xcolor}

\newcommand{\pT}{p_{\rm T}}
\newcommand{\pTjet}{p_{\rm T}^{\rm ch,jet}}
\newcommand{\sqrts}{\sqrt{s}}
\newcommand{\Lc}{\Lambda_{\rm c}^{+}}
\newcommand{\Dz}{{\rm D}^{0}}
\newcommand{\Lz}{\Lambda^{0}}
\newcommand{\Kzs}{{\rm K}^{0}_{\rm S}}
\newcommand{\jT}{j_{\rm T}}
\newcommand{\kT}{k_{\rm T}}
\newcommand{\PYTHIA}{\textsc{Pythia\,8}}
\newcommand{\pp}{$pp$}

\begin{document}
	
	\title[Baryon-to-meson ratios in the Lund jet plane]
	{Baryon-to-meson ratios in the Lund jet plane: resolving
		string-junction and thermal-fragmentation effects}
	
	\author{R\'obert V\'ertesi}
	\address{HUN-REN Wigner Research Centre for Physics,
		P.O.\ Box~49, H-1525 Budapest, Hungary}
	\ead{vertesi.robert@wigner.hu}
	
	\begin{abstract}
		Baryon-to-meson ratios inside charged jets are studied in the
		Lund jet plane using \PYTHIA{} simulations of \pp{} collisions
		at $\sqrts = 14$~TeV, with $15 < \pTjet < 30$~GeV/$c$ in an
		acceptance compatible with the future ALICE\,3 detector.
		Jets are reclustered with the Cambridge--Aachen algorithm and
		identified hadrons are tagged at the splitting where they first
		separate from the leading branch.
		Three model configurations are compared: Monash, color
		reconnection beyond leading color (CR-BLC), and CR-BLC combined
		with thermodynamical string fragmentation (Thermal\,+\,CR-BLC).
		The $\Lc/\Dz$ ratio rises with $\log(1/z)$ in the two
		configurations that include color-string junctions and is
		approximately flat under Monash. The same trend is present at
		the first Cambridge--Aachen branching, indicating that the
		effect is set at the level of the color topology rather than
		through the cumulative action of many soft splittings.
		Light- and strange-sector ratios show no $z$-dependent
		signature of either mechanism, but the angular projection
		reveals non-trivial structure at the first branching that is
		washed out when emissions are summed.
		The Cambridge--Aachen declustering and the Lund plane together
		separate the kinematic signatures of the two
		baryon-enhancement mechanisms in \PYTHIA{} and provide
		observables accessible to future ALICE\,3 measurements.
	\end{abstract}
	
	
	\submitto{\JPG}
	
	\maketitle
	
	\section{Introduction}
	\label{sec:intro}
	
	Baryon-to-meson ratios at intermediate transverse momentum are
	among the most discussed observables in high-energy hadronic
	collisions.
	In heavy-ion collisions, an enhancement of the
	$p/\pi$ and $\Lz/\Kzs$ ratios at intermediate $\pT$
	has long been associated with the interplay between collective
	radial flow and quark recombination in the
	quark--gluon plasma~\cite{Fries:2005}.
	The observation of a similar pattern by the ALICE experiment in
	small systems~\cite{ALICE:journey} has motivated a critical
	reassessment of the underlying mechanisms, since the formation of
	a thermalized medium in \pp{} collisions remains uncertain.
	A central question that has emerged is therefore whether the
	small-system baryon enhancement reflects genuine collective
	final-state effects, or whether it can be accounted for by
	non-perturbative QCD dynamics inherent to the hadronization
	process itself.
	Resolving this question requires observables that probe
	hadronization at a more differential level than the inclusive
	transverse momentum spectra in which the enhancement was
	originally observed.
	
	A natural framework for such a differential analysis is provided
	by jet substructure.
	A high-energy jet is built from a sequence of QCD splittings
	that span a wide range of angular and transverse-momentum
	scales, from the hard scattering down to the non-perturbative
	hadronization region.
	Reclustering the jet with the angular-ordered Cambridge--Aachen
	(C/A) algorithm~\cite{Dokshitzer:1997} unwinds this branching history
	splitting by splitting.
	The Lund jet plane~\cite{Dreyer:2018} maps each branching onto
	the two-dimensional space of angular ($\theta$) and relative-transverse-momentum ($\kT$) scales, providing
	a kinematically uniform representation of the jet's internal
	radiation pattern.
	The same C/A declustering also yields the splitting fraction
	$z$ and the splitting angle $\theta$ as complementary
	one-dimensional projections of the jet substructure.
	Beyond the perturbative regime in which it was originally
	formulated, the Lund plane has become a standard tool for
	quantifying both the perturbative parton shower and the
	non-perturbative effects that modify
	it~\cite{ATLAS:Lund2020,Dreyer:2022}.
	By tagging an identified hadron at the C/A step where it first
	separates from the leading branch, one can ask where, in
	$(\theta,\kT,z)$, a given hadron species is preferentially
	emitted.
	
	To make quantitative predictions for these observables, we
	use the \PYTHIA{} Monte Carlo
	event generator~\cite{Sjostrand:2015,Bierlich:2022pfr},
	which provides a controlled framework in which different
	hadronization mechanisms can be turned on and off independently.
	Two such mechanisms are particularly relevant for baryon
	production. Color reconnection beyond leading color
	(CR-BLC)~\cite{Christiansen:2015} introduces SU(3)-consistent
	string-length minimization and allows for the formation of
	color-string junctions, which act as sources of baryon
	number~\cite{Altmann:2024}.
	The CR-BLC mode~2 tune is known to reproduce the multiplicity
	dependence of charm baryon-to-meson ratios observed by
	ALICE in minimum-bias \pp{} collisions~\cite{ALICE:Lc2022},
	as well as the intra-jet longitudinal momentum fraction
	distributions of $\Lc$ baryons measured in
	Ref.~\cite{ALICE:LcZ2024}.
	Thermodynamical string fragmentation~\cite{Fischer:2017},
	in turn, replaces the Gaussian transverse-momentum suppression
	of the standard Lund model with a Boltzmann-like form
	characterized by an effective temperature. Combined
	with a close-packing mechanism for overlapping strings, it
	enhances the production of strange and multi-strange baryons.
	We refer to the combination of these features as the
	\emph{thermodynamical fragmentation} model in the following.
	
	We have recently shown that intra-jet baryon-to-meson ratios
	provide a sensitive probe of these
	mechanisms~\cite{Vertesi:2025,Ortiz:2026qmf}.
	The $\Lc/\Dz$ enhancement at low $\jT$ inside jets was found
	to be driven by color-string junctions, whereas the light- and
	strange-sector enhancement was associated with the
	thermodynamical fragmentation model. The multiplicity
	dependence was shown to be largely a consequence of a
	quark-to-gluon jet composition shift.
	These analyses, performed in the longitudinal momentum fraction
	$z^{\rm ch}_{\parallel}$ and in the transverse component $\jT$,
	do not however resolve at which stage of the jet evolution
	the enhancement is generated.
	The Lund plane representation introduced above is precisely
	the differential observable needed to address this question.
	
	The analysis is performed in the kinematic regime accessible
	to the ALICE\,3 detector~\cite{ALICE3:LoI}, which is expected
	to provide hadron identification down to low $\pT$ over an
	extended pseudorapidity range, making intra-jet
	identified-hadron measurements feasible at LHC Run~5.
		We note that the present study is carried
		out entirely within the \PYTHIA{} framework, therefore the identification
		of the color-string junction mechanism as the origin of the
		observed signatures is a statement depending on
		this generator. This is discussed further in
		Sec.~\ref{sec:discussion}.
	
	The paper is organized as follows.
	Section~\ref{sec:method} describes the simulation setup and
	the Lund plane tagging procedure.
	Section~\ref{sec:validation} compares the three tunes against
	existing ALICE data on intra-jet observables that have been
	studied in our previous work, in the corresponding kinematic
	regime and simulation setup.
	Sections~\ref{sec:results_z}--\ref{sec:results_2d} present the
	splitting-fraction, angular, and two-dimensional Lund plane
	results, respectively.
	A discussion of the results and their experimental outlook is
	given in Sec.~\ref{sec:discussion}, and conclusions are
	summarized in Sec.~\ref{sec:conclusions}.
	
	\section{Simulation and analysis method}
	\label{sec:method}
	
	Proton--proton collisions are simulated with \PYTHIA{} version
	8.309~\cite{Bierlich:2022pfr} using hard-QCD processes.
	Three model configurations are compared.
	The \textbf{Monash} configuration uses the Monash 2013
	tune~\cite{Skands:2014}, with the standard Lund fragmentation
	function and the MPI-based color-reconnection model.
	The \textbf{CR-BLC} configuration is the CR-BLC mode~2 tune of
	Ref.~\cite{Christiansen:2015}, which adds color-string
	junctions through QCD-consistent string-length minimization
	with strict causality requirements.
	The \textbf{Thermal\,+\,CR-BLC} configuration combines
	thermodynamical string fragmentation~\cite{Fischer:2017} with
	CR-BLC mode~2 and the close-packing mechanism.
	The relevant \PYTHIA{} parameters for each configuration are
	listed in Table~I of Ref.~\cite{Vertesi:2025}.
	
	The reference observables shown in
	Sec.~\ref{sec:validation} are computed at $\sqrts = 13$~TeV in
	charged jets with $\pTjet > 10$~GeV/$c$ and with multiparton
	interactions (MPI) switched off, in order to match the kinematic
	selection of the ALICE measurements~\cite{ALICE:2022ecr,ALICE:Lc2022}
	that they are compared to, and to be consistent with the
	analysis setups of Refs.~\cite{Vertesi:2025,Ortiz:2026qmf}.
	The Lund plane analysis presented in
	Secs.~\ref{sec:results_z}--\ref{sec:results_2d} is performed at
	the LHC Run~5 energy of $\sqrts = 14$~TeV, with charged jets
	selected in the range $15 < \pTjet < 30$~GeV/$c$, and with MPI
	enabled so that the simulated events correspond to the realistic
	\pp{} environment that will be measured by ALICE\,3.
	A total of $10^{8}$ events are generated for each configuration
	in both of the setups.
	
	Jets are reconstructed from final-state charged particles using
	the anti-$\kT$ algorithm~\cite{Cacciari:2008} with resolution
	parameter $R = 0.4$, as implemented in
	FastJet~\cite{Cacciari:2012}, with the $E$-recombination scheme.
	Tracks are required to have $|\eta| < 1.5$ and
	$\pT > 0.05$~GeV/$c$.
	The pseudorapidity range restricts the analysis to the central pseudorapidity region, while fully contained in the	acceptance of ALICE\,3~\cite{ALICE3:LoI}.
	The transverse-momentum threshold is conservative relative to
	the projected ALICE\,3 capability, which is expected to extend
	down to $\pT \sim 20$~MeV/$c$ for charged-pion identification.
	The effect of a lower threshold on the Lund plane reach is discussed in Sec.~\ref{sec:discussion}.
	Charged jets are required to have $|\eta_{\rm jet}| < \eta_{\rm jet}^{\rm max}$,
	with $\eta_{\rm jet}^{\rm max} = 1.5 - R = 1.1$, ensuring full containment of
	the jet area within the tracking acceptance.
	
	Each selected jet is reclustered with the Cambridge--Aachen
	algorithm~\cite{Dokshitzer:1997} with the same resolution
	parameter, and the resulting clustering tree is unwound from
	the root toward the leaves.
	Because C/A is angular-ordered, the root-level splitting is
	the widest-angle branching of the jet.
	At each declustering step, the jet $j$ splits into two subjets
	$j_1$ and $j_2$, with $\pT^{j_1} \geq \pT^{j_2}$.
	The Lund plane coordinates of the splitting are
	\begin{equation}
		x = \ln\!\left(\frac{R}{\Delta R_{12}}\right), \qquad
		y = \ln\!\left(\frac{\kT}{\mathrm{GeV}}\right),
	\end{equation}
	with $\Delta R_{12}$ the rapidity--azimuth distance between the
	subjets and $\kT = \pT^{j_2}\,\Delta R_{12}$ the relative
	transverse momentum of the softer subjet.
	The subscript on $\Delta R_{12}$ is omitted in figure axes when
	no ambiguity arises.
	The splitting fraction and angle are
	$z = \pT^{j_2}/(\pT^{j_1}+\pT^{j_2})$ and
	$\theta = \Delta R_{12}/R$.
	After each step the harder subjet $j_1$ is followed for the
	next declustering.
	
	For each jet we examine identified $\pi$, ${\rm K}$, $p$, $\Kzs$, $\Lz$, $\Xi$, $\Omega$, $\Dz$, and $\Lambda_{\rm c}^{+}$ hadrons.
	Throughout this paper, charge conjugates are implicitly included
	in each species, so that for example $\Lambda_{\rm c}^{+}$
	stands for $\Lambda_{\rm c}^{+} + \Lambda_{\rm c}^{-}$ and $\Dz$
	for $\Dz + \overline{\rm D}{}^{0}$.
	As the C/A tree is walked from the root toward the leaves, a hadron initially follows the harder subjet at each step until, at some splitting, the harder subjet no longer contains it and the hadron ends up in the softer subjet $j_2$. This splitting is recorded as the one at which the hadron tags the emission, and the corresponding splitting variables $z,\theta$ and $\kT$ are stored.
	Each hadron therefore tags exactly one splitting per jet, at which it parts from the leading branch. In addition, we keep separate histograms restricted to the widest-angle (root-level) splitting of each jet, in order to study the primary branching independently of the rest of the clustering tree.
	
	Feed-down is included in each species. For instance, a $\Lc$ that decays into a $\Lz$ contributes to both the $\Lc$ and $\Lz$ histograms. Using this approach the results match pre-feed-down-correction experimental observables.
	
	The tagged-emission histograms are normalized to the total number
	of accepted jets, $N_{\rm jets}$.
	Ratios of per-jet yields between species are then formed. Bins
	in which the relative statistical uncertainty exceeds 50\,\% are
	omitted from the figures.
	
	\section{Reference observables}
	\label{sec:validation}
	
\begin{figure}[t]
		\centering
		\includegraphics[width=0.9\linewidth]{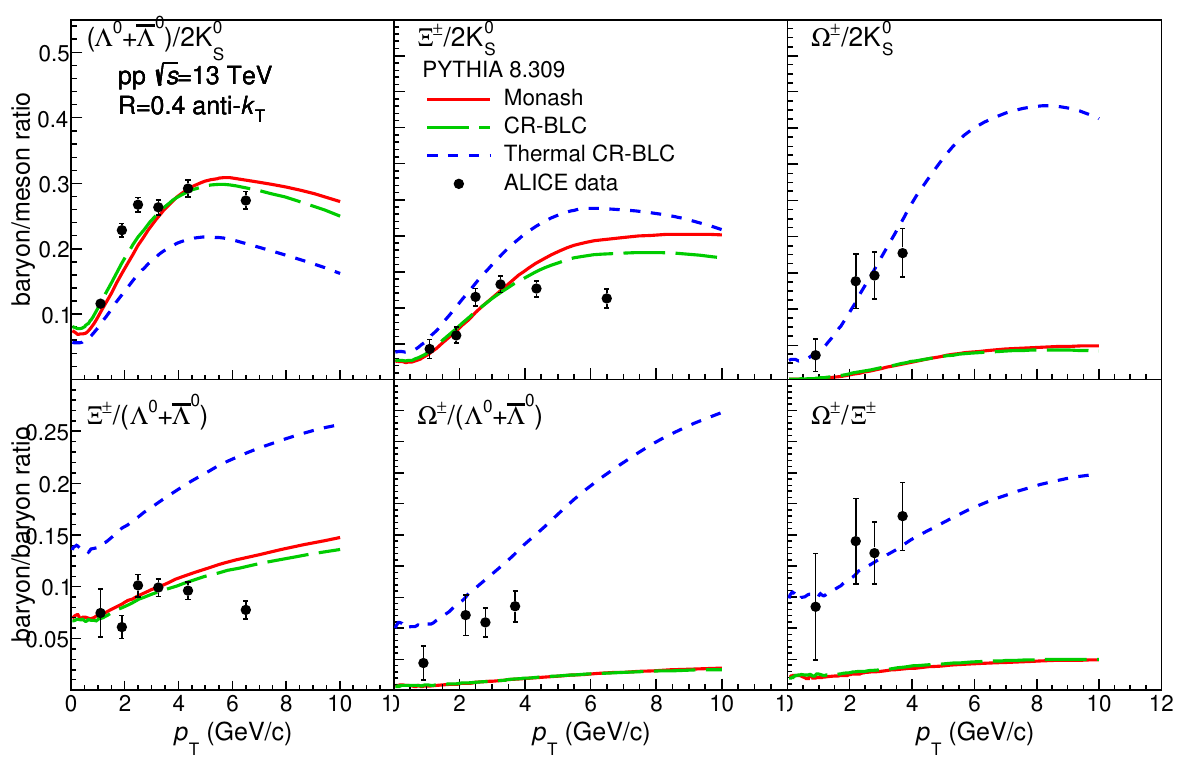}
		\caption{Intra-jet baryon-to-meson and baryon-to-baryon
			ratios for the strange and multi-strange sectors as
			a function of constituent $\pT$ in charged jets with
			$\pTjet > 10$~GeV/$c$ in \pp{} collisions at
			$\sqrts = 13$~TeV, simulated with MPI off, for the
			three \PYTHIA{} configurations described in the text.
			ALICE data are from Ref.~\cite{ALICE:2022ecr}. \textsc{Pythia} 8.309 is used throughout.}
		\label{fig:validation_strange}
	\end{figure}

	Before examining the Lund plane, we briefly compare the three
	tunes for intra-jet baryon-to-meson ratios as a function of
	constituent $\pT$, in the simulation setup matched to the
	existing ALICE measurements
	($\sqrts = 13$~TeV, $\pTjet > 10$~GeV/$c$, MPI off,
	see Sec.~\ref{sec:method}).
	ALICE has reported strange and multi-strange baryon-to-meson
	and baryon-to-baryon ratios inside and outside of
	jets~\cite{ALICE:2022ecr}, shown in
	Fig.~\ref{fig:validation_strange} together with the three model
	predictions.
	Monash and CR-BLC give nearly identical predictions for
	$\Lz/\Kzs$ and $\Xi^{\pm}/\Kzs$, indicating that the addition of
	color-string junctions alone has little effect on light or
	strange baryon production.
	Thermal\,+\,CR-BLC reproduces the qualitative strangeness
	hierarchy but overpredicts the multi-strange yields.
	This is a known limitation of the model that has been
	discussed in Refs.~\cite{Fischer:2017,Vertesi:2025} but that
	does not affect the conclusions of the present analysis,
	which deals with $z$- and $\theta$-dependent shapes rather
	than with absolute multi-strange yields.
	
\begin{figure}[t]
		\centering
		\includegraphics[width=0.5\linewidth]{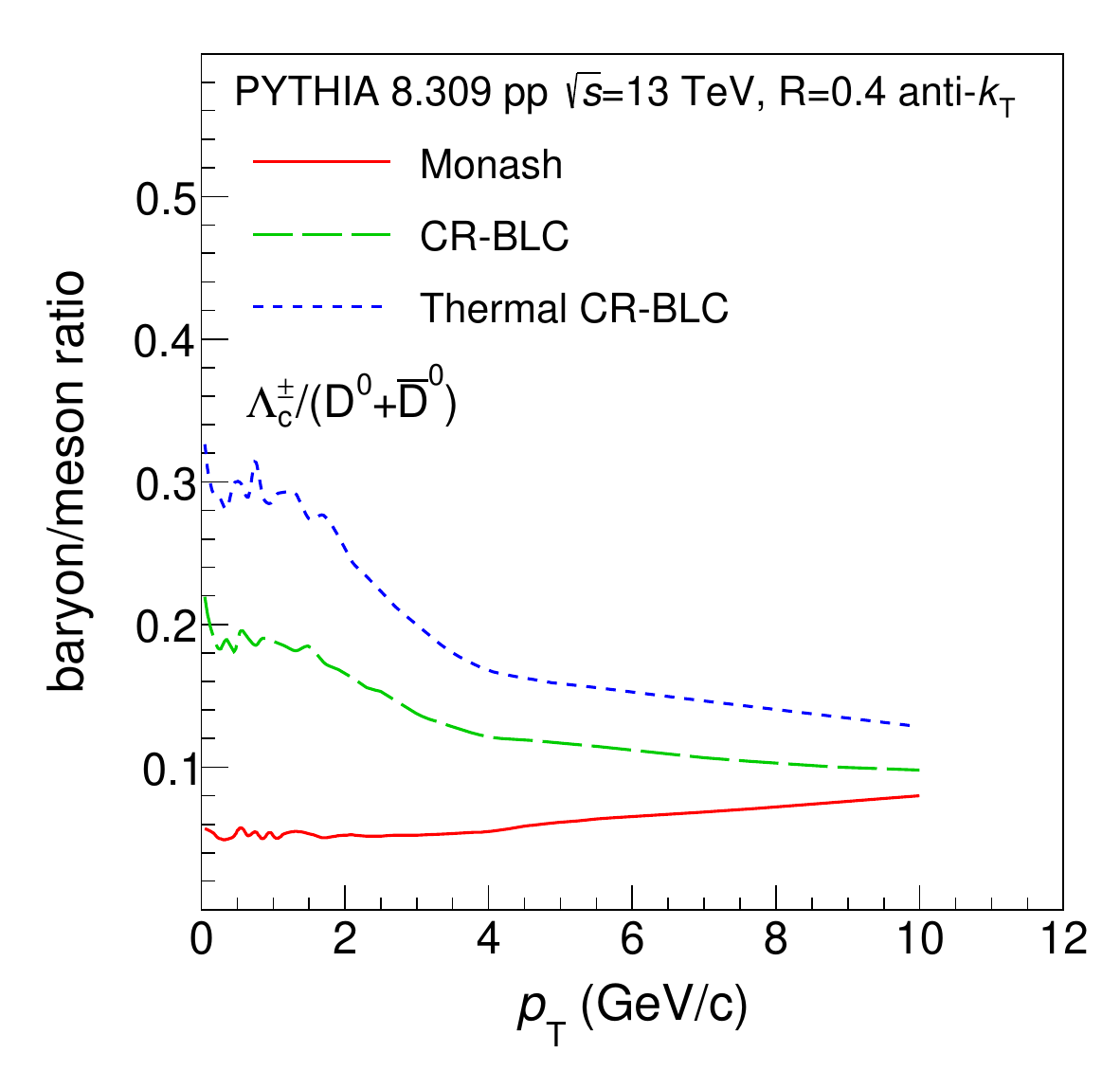}
		\caption{$\Lc/\Dz$ ratio as a function of constituent $\pT$
			inside charged jets with $\pTjet > 10$~GeV/$c$ in
			\pp{} collisions at $\sqrts = 13$~TeV, simulated with \textsc{Pythia} 8.309 and
				MPI off, for the three configurations
			described in the text.}
		\label{fig:validation_charm}
	\end{figure}

	Figure~\ref{fig:validation_charm} shows the
	$\Lc/\Dz$ ratio inside jets in the same simulation setup.
	The three tunes are clearly separated: Monash predicts a small,
	nearly flat ratio of about $0.05$--$0.10$, CR-BLC a larger ratio
	that decreases from $\approx 0.20$ at low constituent $\pT$ to
	$\approx 0.12$ at high $\pT$, and Thermal\,+\,CR-BLC a still
	larger ratio with similar shape.
	The ordering of the three configurations is preserved at all constituent momenta, which motivates the Lund plane analysis below,
	designed to reveal at which jet splittings the
	enhancement is built up.
	
	\section{Splitting-fraction results}
	\label{sec:results_z}
	
	We now detail the analysis of the C/A declustering in the
	kinematic regime defined in Sec.~\ref{sec:method}
	($\sqrts = 14$~TeV, $15 < \pTjet < 30$~GeV/$c$, MPI on).
	We begin with the splitting-fraction projection,
	$z = \pT^{j_2}/(\pT^{j_1}+\pT^{j_2})$, which captures the
	asymmetry of each Cambridge--Aachen branching independently of
	the angular and transverse-momentum scales of the splitting,
	and is a complementary observable to the Lund plane axes.
	We consider both the inclusive sum over all C/A declustering
	steps and the restriction to the first (root-level, widest-angle)
	branching, with no grooming or soft-drop-like
	selection~\cite{Larkoski:2014} applied.
	
\begin{figure}[t]
		\centering
		\includegraphics[width=\linewidth]{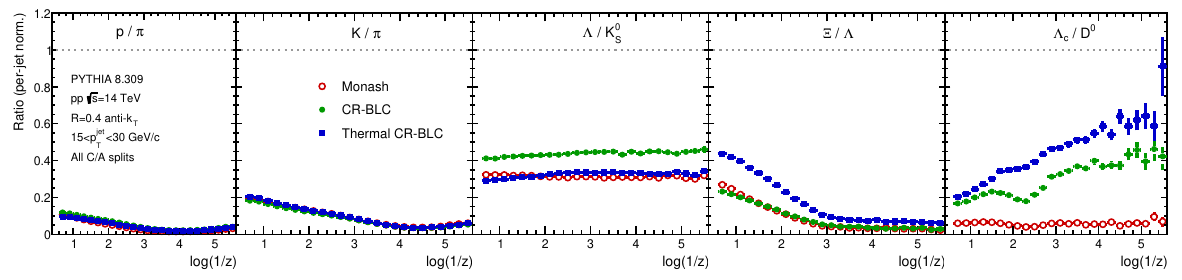}
		\caption{Hadron species ratios as a function of $\log(1/z)$,
			summed over all Cambridge--Aachen declustering steps,
			in charged jets with $15 < \pTjet < 30$~GeV/$c$ in
			\pp{} collisions at $\sqrts = 14$~TeV, simulated
			with MPI on, for the three \textsc{Pythia} 8.309 configurations. 
				Statistical uncertainties are
				shown on all points, however, they become visible mainly in the softest bins.}
		\label{fig:logz_all}
	\end{figure}

	Figure~\ref{fig:logz_all} shows all five species ratios as a
	function of $\log(1/z)$, summed over all C/A declustering steps.
	Histograms entering the ratios are normalized to the per-jet
	yield so that the comparison between tunes is not biased by
	overall production-rate differences.
	
\begin{figure}[t]
		\centering
		\includegraphics[width=\linewidth]{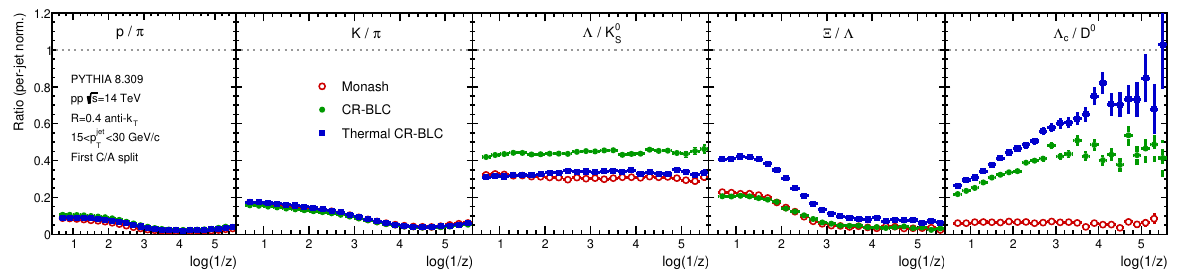}
		\caption{Hadron species ratios as a function of $\log(1/z)$,
			restricted to the first (widest-angle) C/A branching
			of each jet, in charged jets with
			$15 < \pTjet < 30$~GeV/$c$ in \pp{} collisions at
			$\sqrts = 14$~TeV, simulated with MPI on, for the
			three \textsc{Pythia} 8.309 configurations.}
		\label{fig:logz_first}
	\end{figure}

	The most pronounced feature is the $\log(1/z)$ dependence of
	the $\Lc/\Dz$ ratio in the two configurations that include
	string junctions.
	Under CR-BLC, $\Lc/\Dz$ rises from $\approx 0.15$ at hard
	splittings ($\log(1/z) \lesssim 1$) to $\approx 0.4$ at soft
	splittings ($\log(1/z) \gtrsim 5$); under Thermal\,+\,CR-BLC,
	it rises from $\approx 0.2$ to values $\approx 0.6$ in
	the softest accessible bins.
	Under Monash, the ratio is small ($\lesssim 0.1$) and
	approximately flat across the full range.
	The absence of a $\log(1/z)$ trend in Monash, and the appearance
	of a clear trend as soon as junctions are switched on, 
		is consistent with the junction mechanism being the origin of the
		$z$-dependent charm baryon enhancement seen in the C/A
		declustering, within the \PYTHIA{} framework studied here.
	The separation between the CR-BLC and Monash predictions
		for $\Lc/\Dz$ is statistically significant over the full shown range. Computing a bin-by-bin significance
		$S = \frac{|R_{\rm CR\textrm{-}BLC} - R_{\rm Monash}|}{\sqrt{\sigma_{\rm CR\textrm{-}BLC}^{2}+\sigma_{\rm Monash}^{2}}}$ 
		from the per-jet-normalized histograms underlying
		Figs.~\ref{fig:logz_all} and~\ref{fig:logz_first}
		($1.10$--$1.13\times10^{7}$ selected jets per configuration),
		we find $S$ in excess of several tens of standard deviations
		for $\log(1/z) \lesssim 3$--$4$. At the softest bin
		retained by the 50\,\% relative uncertainty cut, $S = 4.4$ for
		the all-emissions selection ($\log(1/z)=5.5$) and $S = 4.0$ for
		the first-split selection ($\log(1/z)=5.3$). The Monash and
		CR-BLC predictions are therefore never separated by less than
		about $4\sigma$ anywhere in the plotted range.
	The thermal weights of Thermal\,+\,CR-BLC act on top of the
	junction effect: they shift the absolute level of $\Lc/\Dz$
	upward by roughly a factor of 1.5 relative to CR-BLC, while
	preserving the rising shape (see Sec.~\ref{sec:discussion} for
	an interpretation).
	
	The light- and strange-sector ratios behave very differently
	from the charm sector.
	The $p/\pi$ and ${\rm K}/\pi$ ratios decrease monotonically
	with $\log(1/z)$ for all three tunes, with the three model
	predictions essentially overlapping.
	This monotonic decrease is a kinematic consequence of the
	fragmentation function: heavier hadrons (kaons and protons,
	relative to pions) tend to be produced at larger
	splitting fractions and therefore depopulate the soft-$z$ region.
	The $\Xi/\Lz$ ratio similarly decreases with $\log(1/z)$ for
	all three tunes, with Thermal\,+\,CR-BLC being higher than CR-BLC and Monash, reflecting the
	mild enhancement of multi-strange baryons under thermodynamical
	fragmentation.
	The $\Lz/\Kzs$ ratio is approximately flat in $\log(1/z)$ but
	clearly separated in absolute level: CR-BLC sits highest at
	$\approx 0.42$, while Monash and Thermal\,+\,CR-BLC overlap
	at $\approx 0.30$.
	This ordering (CR-BLC above the other two) is the only
	case in our analysis where Monash and CR-BLC differ
	significantly in a light- or strange-sector observable, and
	indicates that color-string junctions do enhance light strange
	baryon production at the level of the inclusive C/A projection,
	although without introducing a $z$-dependence.
		This level shift, rather than a shape change, can be
		connected to the junction mechanism itself. CR-BLC and Monash are
		not expected to differ through independently tuned light-flavor
		fragmentation parameters, since both build on the same underlying
		Lund fragmentation framework. The difference can more likely be traced to 
		the junction. An SU(3)-consistent junction needs three string legs
		meeting at a point and must eventually fragment into a baryon, and
		this extra fragmentation channel would compete with ordinary
		$q\bar{q}$ string breaks without necessarily changing how momentum
		is shared at any individual break. A natural expectation is
		that its effect is mainly to raise the overall probability of
		producing a baryon like $\Lz$ relative to $\Kzs$, rather than to
		bias the splitting fraction at which it is
		produced~\cite{Christiansen:2015,Altmann:2024}.
	The addition of the thermodynamical fragmentation model on
	top of CR-BLC returns the $\Lz/\Kzs$ ratio to a level
	comparable to Monash, while the $\Xi/\Lz$ ordering
	Thermal\,+\,CR-BLC above CR-BLC and Monash noted above
	indicates that the strange-sector level shifts respond to the
	two mechanisms in a coupled way. A possible interpretation in
	terms of species redistribution is discussed in
	Sec.~\ref{sec:discussion}.
	
	Crucially, none of the light- or strange-sector ratios
	distinguishes Monash and CR-BLC in their $\log(1/z)$
	shape: the differences between these two configurations are
	at most overall normalization shifts, not $z$-dependent trends.
	This is consistent with the picture established in
	Refs.~\cite{Vertesi:2025,Ortiz:2026qmf}: junctions and
	thermodynamical fragmentation modify the absolute yields of
	light and strange baryons but do not reshape their splitting
	kinematics in the C/A tree.
	
	Figure~\ref{fig:logz_first} shows the same ratios restricted
	to the first (widest-angle) C/A branching of each jet.
	The $\Lc/\Dz$ rise with $\log(1/z)$ persists with a shape and
	magnitude very similar to those of the all-emissions plot.
	The pattern is therefore not generated through accumulation over
	many soft splittings deep in the shower, but is already imprinted
	on the primary splitting.
	This is consistent with junctions being formed during the
	color-reconnection phase, prior to hadronization.
	The light- and strange-sector ratios remain qualitatively
	similar in $\log(1/z)$ in the first-emission case relative to
	the all-emissions case: the $p/\pi$, ${\rm K}/\pi$ and
	$\Xi/\Lz$ kinematic decreases are preserved, and the level
	ordering of $\Lz/\Kzs$ (CR-BLC above Monash and
	Thermal\,+\,CR-BLC) is unchanged.
	This stability of the $z$-projection under restriction to the
	primary branching is in marked contrast to the angular
	projection discussed in the next section.
	
	\section{Angular dependence}
	\label{sec:results_theta}
	
\begin{figure}[t]
		\centering
		\includegraphics[width=\linewidth]{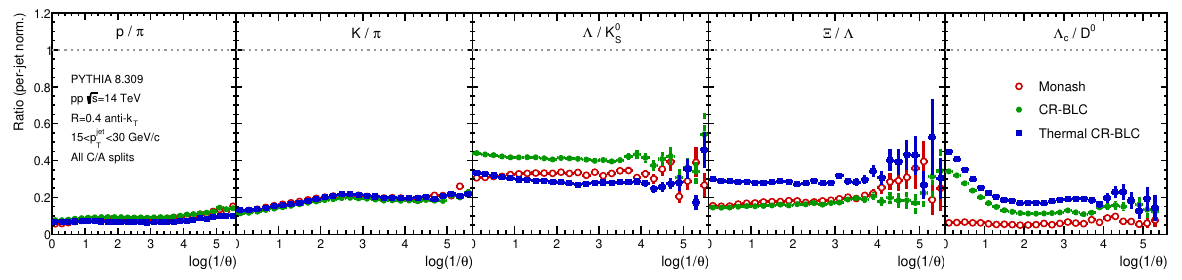}
		\caption{Hadron species ratios as a function of
			$\log(1/\theta)$, summed over all Cambridge--Aachen
			declustering steps, in charged jets with
			$15 < \pTjet < 30$~GeV/$c$ in \pp{} collisions at
			$\sqrts = 14$~TeV, simulated with MPI on, for the
			three \textsc{Pythia} 8.309 configurations.}
		\label{fig:logtheta_all}
	\end{figure}

\begin{figure}[t]
		\centering
		\includegraphics[width=\linewidth]{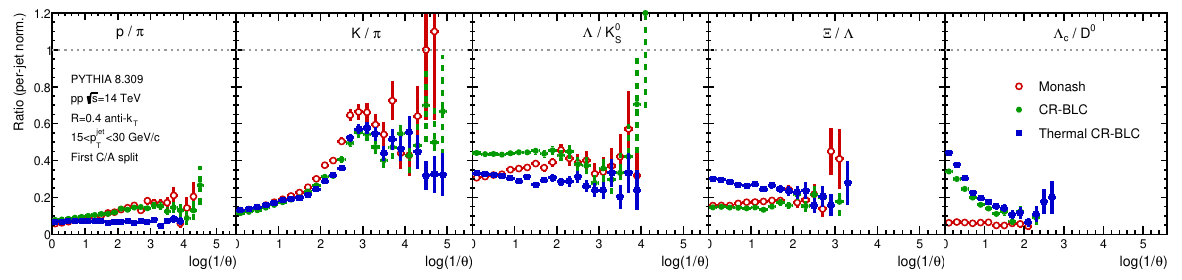}
		\caption{Hadron species ratios as a function of
			$\log(1/\theta)$ at the first (widest-angle) C/A
			branching of each jet, in charged jets with
			$15 < \pTjet < 30$~GeV/$c$ in \pp{} collisions at
			$\sqrts = 14$~TeV, simulated with MPI on, for the
			three \textsc{Pythia} 8.309 configurations.}
		\label{fig:logtheta_first}
	\end{figure}

	The angular dependence of the same ratios is shown in
	Fig.~\ref{fig:logtheta_all} summed over all C/A declustering
	steps, and Fig.~\ref{fig:logtheta_first} restricted to the
	first branching.
	A direct comparison between the two figures provides a
	qualitatively different picture from the splitting-fraction
	projection: while the $z$-projection is essentially preserved
	when going from all emissions to the first split, the angular
	projection exhibits substantially more structure at the first
	branching that is largely washed out when emissions are summed.
	
	Considering first the all-emissions projection
	(Fig.~\ref{fig:logtheta_all}), the light-sector ratios
	$p/\pi$ and ${\rm K}/\pi$ rise monotonically and smoothly with
	$\log(1/\theta)$ for all three tunes, with the predictions
	essentially overlapping.
	The strange-sector ratios are predominantly flat and separated
	in level: $\Lz/\Kzs$ shows the same CR-BLC above Monash and
	Thermal\,+\,CR-BLC ordering as in the $z$-projection,
	$\Xi/\Lz$ shows the Thermal\,+\,CR-BLC above CR-BLC and
	Monash ordering, and both cases reflect the strange-sector
	level shifts discussed in Sec.~\ref{sec:results_z}.
	The $\Lc/\Dz$ ratio decreases monotonically with
	$\log(1/\theta)$ for all three configurations: starting from
	$\approx 0.4$ at large angles, CR-BLC and Thermal\,+\,CR-BLC
	fall to $\approx 0.15$ at the smallest accessible angles, while
	Monash decreases from $\approx 0.20$ to $\approx 0.05$.
	The angular trend is opposite in sign to the $z$-trend, where
	$\Lc/\Dz$ rises with $\log(1/z)$, and the offset between
	junction-active and Monash configurations is preserved across
	the full angular range.
	
	At the primary branching (Fig.~\ref{fig:logtheta_first}), several non-trivial features emerge that are not present in the all-emissions case.
	    The light-sector $p/\pi$ ratio rises smoothly with
		$\log(1/\theta)$ for both Monash and CR-BLC, from
		$\approx 0.06$ (Monash) and $\approx 0.08$ (CR-BLC) at
		$\log(1/\theta) \approx 0.1$ ($\theta \approx 0.9\,R$) to
		$\approx 0.15$--$0.17$ by $\log(1/\theta) \approx 3.5$
		($\theta \approx 0.03\,R$), beyond which the per-bin
		relative statistical uncertainty exceeds $\approx 20\,\%$. The
		two tunes essentially overlap each other within these uncertainties over the
		full range. We do not find statistically significant evidence
		for a non-monotonic behavior in	this ratio for either tune. In contrast, the
	${\rm K}/\pi$ ratio exhibits a pronounced peak around
	$\log(1/\theta) \approx 3$, corresponding to splitting angles
	$\theta \sim 0.05\,R$, present for all three tunes with
	different amplitudes (Monash highest, Thermal\,+\,CR-BLC
	lowest). The position of the peak is essentially tune-independent
	and reflects the typical angular scale at which kaons emerge
	from the primary branching at this jet $\pT$, while its amplitude 
	reflects the overall ${\rm K}/\pi$ yield ratio in each tune.
	In the strange sector, $\Lz/\Kzs$ separates the three tunes
	both in level and in shape: CR-BLC is highest with a slight
	rise toward small angles, Monash is intermediate and mildly
	decreasing, and Thermal\,+\,CR-BLC is lowest and approximately
	flat.
	The $\Xi/\Lz$ ratio decreases with $\log(1/\theta)$ for 
	Thermal\,+\,CR-BLC, with its elevated overall level noted earlier; the Monash and CR-BLC slopes are negligible.
	The $\Lc/\Dz$ ratio shows a mild decrease with $\log(1/\theta)$
	in CR-BLC and Thermal\,+\,CR-BLC, with similar magnitudes as
	in the all-emissions case.
	
	The first-split and all-emissions projections carry complementary information.
	The peak structure in ${\rm K}/\pi$ and the differentiated shape
	of $\Lz/\Kzs$ at the first branching are both washed out when
	secondary splittings are added.
	    We show in
		Sec.~\ref{sec:discussion} that this washing-out is a shape
		effect of the secondary-splitting contribution, rather than a
		difference in the overall yield fraction contributed by
		secondary splittings between mesons and baryons.
	This is in contrast to the splitting-fraction projection, where
	secondary emissions reproduce the soft-$z$ behavior of the
	primary branching and the all-emissions and first-split plots
	are nearly identical.
	The first branching is therefore the cleaner observable for
	studying angular structure, while either projection is
	acceptable for studying $z$-dependence.
	
	\section{Two-dimensional Lund plane}
	\label{sec:results_2d}
	
\begin{figure}[t]
		\centering
		\includegraphics[width=\linewidth]{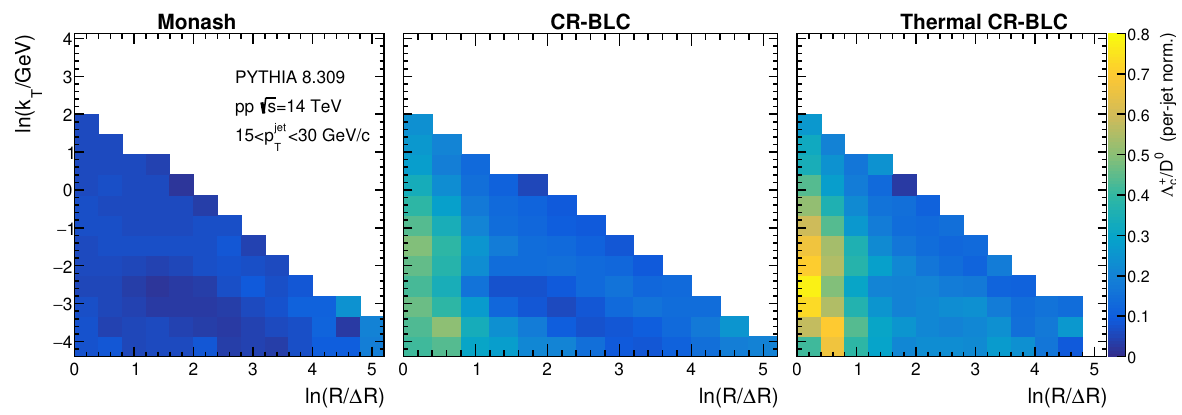}
		\caption{$\Lc/\Dz$ ratio in the two-dimensional Lund plane
			$[\ln(R/\Delta R),\,\ln(\kT/\mathrm{GeV})]$ for the
			three \textsc{Pythia} 8.309 configurations, in charged jets with
			$15 < \pTjet < 30$~GeV/$c$ in \pp{} collisions at
			$\sqrts = 14$~TeV, simulated with MPI on.}
		\label{fig:lund2d}
	\end{figure}

	Figure~\ref{fig:lund2d} shows the $\Lc/\Dz$ ratio in the
	Lund plane for the three tunes.
	The numerator and denominator are independently normalized to
	the per-jet yield prior to the division, and a $4\times 4$
	rebinning is applied to reduce the statistical fluctuations
	that would otherwise dominate the charm sector.
	
	Under Monash, the ratio is small and approximately uniform over
	the populated region of the plane.
	Under CR-BLC, the ratio is enhanced everywhere relative to
	Monash, with a tendency to grow toward small $\kT$ at fixed
	angle, mirroring the $\log(1/z)$ trend of
	Fig.~\ref{fig:logz_all}.
	Under Thermal\,+\,CR-BLC, the enhancement is the strongest of
	the three configurations, with the largest values in the soft
	($\ln(\kT/\mathrm{GeV}) \lesssim 0$) and wide-angle (small
	$\ln(R/\Delta R)$) corner of the plane.
	The two-dimensional pattern is consistent with the
	one-dimensional variables discussed above:
	the $\kT$ axis captures the rising trend of $\Lc/\Dz$ with
	decreasing $\kT$ that drives the $\log(1/z)$ enhancement, while
	the angular axis captures the falling trend of $\Lc/\Dz$ with
	$\log(1/\theta)$.
	The two trends combine to localize the junction-induced charm
	baryon enhancement in the soft, wide-angle region of the Lund
	plane -- precisely the region where color-string junctions
	operate kinematically.
	Light- and strange-sector ratios in the same plane representation
	exhibit no comparable structure and are not shown.
	
	\section{Discussion}
	\label{sec:discussion}
	
	The C/A declustering of the jet, together with its projection
	onto the Lund jet plane, offers a kinematically differential
	handle on intra-jet baryon production that complements the
	$\jT$ and $z^{\rm ch}_{\parallel}$ observables discussed in our
	previous work~\cite{Vertesi:2025,Ortiz:2026qmf}.
	Two structural features of the results are worth highlighting
	before turning to the experimental outlook.
	
	The first is the asymmetry between the $z$- and $\theta$-projections
	under restriction to the primary branching.
	Non-trivial angular structure in the light and strange sectors --
	notably a ${\rm K}/\pi$ peak at $\log(1/\theta) \approx 3$ and a
	differentiated $\Lz/\Kzs$ shape -- is visible at the first C/A
	branching and is washed out when emissions are summed, while the
	$z$-projection is essentially preserved between the two cases.
	This follows from the kinematics of the C/A tree. 
		A uniform reshaping of the
		small-$\theta$ region by secondary emissions would leave species
		ratios unchanged if it acted identically on numerator and
		denominator, so the washing-out requires the secondary splitting
		contribution to differ in shape between numerator and denominator
		species. We checked whether this arises from secondary splittings
		contributing a different fraction of the total yield for
		mesons than for baryons, and found no such asymmetry: splittings
		beyond the first account for $\approx 69$--$77\,\%$ of the
		all-emissions yield for all the particles $\pi$, ${\rm K}$, $\Lz$ and $\Lc$,
		with no meson/baryon separation. The washing-out is instead a
		shape effect: isolating the secondary-splittings-only
		contribution (the difference between the all-emissions and
		first-split histograms of Figs.~\ref{fig:logtheta_all}
		and~\ref{fig:logtheta_first}) shows that its ${\rm K}/\pi$
		ratio is comparatively flat, $\approx 0.18$--$0.25$ over
		$\log(1/\theta) \approx 2$--$6$ in all three tunes, in contrast
		to the pronounced first-split-only peak of
		$\approx 0.55$--$0.66$ at $\log(1/\theta) \approx 2.9$--$3.1$.
		Because secondary splittings supply the majority of the $\pi$
		and ${\rm K}$ yield, this flat component dominates the
		yield-weighted sum entering the all-emissions ratio and erases
		the peak, even though the overall yield fractions contributed by
		primary and secondary splittings do not differ between the two
		meson species. This contrasts with the $\Lc/\Dz$ $\log(1/z)$
		trend (Sec.~\ref{sec:results_z}), which persists between the
		first-split and all-emissions selections.
	The soft-$z$ region, by contrast, is populated similarly by
	primary and secondary emissions.
	The primary branching is therefore the cleaner observable for
	angular studies of jet substructure, while either projection
	is acceptable for studying $z$-dependence.
	
	The second is the contrasting interplay of the junction mechanism
	and thermodynamical fragmentation in the charm and strange sectors.
	In the charm sector, the thermal weights raise $\Lc/\Dz$ by an
	approximately $z$-independent factor of $\sim 1.5$ while preserving
	the shape of the $\log(1/z)$ rise, consistent with the two
	mechanisms acting largely independently.
	In the strange sector they instead compensate the junction-induced
	enhancement of $\Lz/\Kzs$ and simultaneously raise $\Xi/\Lz$,
	pointing to a species-redistribution effect of the Boltzmann
	weights among strange baryons: the available strange-baryon
	yield is reshuffled among the $\Lz$, $\Xi$ and $\Omega$ species
	rather than uniformly amplified.
	The same mechanisms thus combine additively in the charm sector
	and redistributively in the strange sector, a distinction that
	is itself a testable feature of the model.
	
	We have also verified that the qualitative pattern -- a
	$\log(1/z)$ rise of $\Lc/\Dz$ in CR-BLC and Thermal\,+\,CR-BLC,
	and flat behavior in the light and strange sectors -- persists
	at much higher jet transverse momenta, in a setup matched to the
	CMS high-$\pTjet$ jet measurements (full jets with $R=0.8$ and
	$550 < \pTjet < 1000$~GeV/$c$, see Ref.~\cite{Vertesi:2025}).
	The magnitude of the $\Lc/\Dz$ rise is however significantly
	reduced at the harder scales, to about a factor of two over the
	accessible $\log(1/z)$ range, compared with a factor of roughly
	five at the ALICE\,3 jet $\pT$.
	This is consistent with junction effects being most prominent at
	soft non-perturbative scales, and identifies the low-$\pTjet$
	regime studied here as the kinematic window with the largest
	sensitivity to the junction mechanism in the Lund plane.
	
	The size of the predicted Monash vs.\ CR-BLC separation in
	$\Lc/\Dz$ at large $\log(1/z)$ suggests that even a modest first-look
	measurement should be able to discriminate between
	junction-based and junction-less hadronization pictures inside
	jets.
	    As quantified in Sec.~\ref{sec:results_z}, this separation
		exceeds several tens of standard deviations for
		$\log(1/z) \lesssim 3$--$4$ and remains above $\approx 4\sigma$
		throughout the plotted range at our current MC statistics
		($1.10$--$1.13\times10^{7}$ selected jets per configuration).
		Rescaling the statistical errors shows that the separation would
		remain above $3\sigma$ in the most marginal (softest) bin even
		with roughly half of this generated jet sample. Translating this
		MC-level statement into a concrete projected ALICE\,3 Run~5
		sensitivity requires the expected $\Lc$ and $\Dz$ reconstruction
		efficiency and purity in this jet-$\pT$ and hadron-$\pT$ range,
		which are beyond the scope of the present phenomenological study.
	Reconstructing $\Lc$ and $\Dz$ inside low-$\pT$ jets requires
	both excellent secondary-vertex resolution and large statistics,
	both of which are central design drivers of the ALICE\,3
	detector~\cite{ALICE3:LoI}.
	A further opportunity is offered by the projected ALICE\,3
	soft-track tracking capability, which is expected to extend the
	charged-track reach down to $\pT \sim 20$~MeV/$c$, below the
	conservative threshold of $0.05$~GeV/$c$ used in the present
	analysis.
	Lowering the track $\pT$ threshold extends the populated region
	of the Lund plane toward the soft direction, and increases the
	accessible range of $\log(1/z)$ correspondingly, into the region
	where the experimental discrimination power of the proposed
	measurement is largest.
	Mitigation of the underlying-event contribution to soft tracks
	will be important for a clean reconstruction of the soft Lund
	plane region, and is a natural target for grooming techniques
	such as soft drop~\cite{Larkoski:2014}.
	
	Several limitations of the present study should be kept in mind.
	    The most important of these is that the analysis is
		performed entirely within \PYTHIA{}. The identification of the
		color-string-junction mechanism as the origin of the observed
		$z$-dependence of $\Lc/\Dz$ is therefore a statement about the
		\PYTHIA{} modeling framework, not a model-independent conclusion,
		and a different hadronization mechanism could in principle
		produce a qualitatively similar soft, wide-angle enhancement.
		Cluster-fragmentation generators such as
		Herwig~\cite{Bellm:2015jjp} do not implement string junctions and
		enhance baryon production through a different mechanism (cluster
		splitting and decay kinematics), alternative junction
		implementations~\cite{Altmann:2024} could shift the quantitative
		$\log(1/z)$ onset scale, and dynamic charm-hadronization
		models~\cite{Bierlich:2024} connect charm enhancement to the local
		string environment rather than to a global junction topology.
		Distinguishing between these pictures requires a dedicated
		cross-generator comparison, which is beyond the scope of the
		present paper and a natural direction for future work.
	The charm sector statistics are limited at the largest values of
	$\log(1/z)$, where the rising trend is most pronounced; the
	conclusions are robust against the 50\,\% relative-uncertainty
	cut applied to the figures, but a higher statistical reach would
	be desirable, in particular for the two-dimensional Lund plane.
	Multiparton interactions are kept enabled in the Lund plane
	analysis, so that the underlying event contributes to the jet
	constituents.
	At the relatively low jet transverse momenta studied here, the
	underlying-event contribution is non-negligible, and a
	multiplicity-differential analysis -- as performed in
	Ref.~\cite{Vertesi:2025} -- would be needed to disentangle it
	from the genuine intra-jet effects.
	We have verified, however, that the qualitative pattern of the
	$\Lc/\Dz$ rise with $\log(1/z)$ is generated by the in-jet C/A
	declustering itself, since the same trend is present in the
	first-emission sample where soft underlying-event contributions
	are kinematically suppressed by the wide-angle requirement.
	
	\section{Conclusions}
	\label{sec:conclusions}
	
	We have studied baryon-to-meson ratios inside charged jets in
	the Lund plane representation, using \PYTHIA{} simulations of
	\pp{} collisions at $\sqrts = 14$~TeV with three model
	configurations: Monash, CR-BLC, and Thermal\,+\,CR-BLC.
	
	In the charm sector, the $\Lc/\Dz$ ratio rises with $\log(1/z)$
	in CR-BLC and in Thermal\,+\,CR-BLC, but is approximately flat
	under Monash.
	CR-BLC differs from Monash through QCD-consistent color
	reconnection, whose main baryon-number-carrying ingredient is
	the color-string junction.
	The absence of a $z$-dependent signature in the light sector,
	combined with the appearance of a clear $z$-dependent rise of
	$\Lc/\Dz$ as soon as CR-BLC is switched on, 
	    is consistent
		with the junction mechanism being the origin of the charm-sector
		$z$-dependence within the \PYTHIA{} framework studied here; as
		discussed in Sec.~\ref{sec:discussion}, confirming this
		model-independently will require comparison with generators
		implementing alternative hadronization pictures.
	The same trend is present at the first Cambridge--Aachen
	branching, indicating that the effect is set on the timescale
	of color reconnection rather than through the cumulative action
	of many soft splittings, consistently with the timing of color
	reconnection in the model, which precedes hadronization.
	The corresponding angular projection $\log(1/\theta)$ shows
	$\Lc/\Dz$ decreasing toward small angles in the junction-active
	configurations, opposite in sign to the $z$-trend, and the two
	trends combine to localize the charm baryon enhancement in the
	soft, wide-angle region of the two-dimensional Lund plane.
	
	In the light and strange sectors, none of the splitting-fraction
	ratios exhibits a $z$-dependent signature of either junctions or
	thermodynamical fragmentation; differences between the three
	configurations appear only as overall level shifts.
	The enhancement of $\Lz/\Kzs$ under CR-BLC relative to Monash
	indicates that junctions do contribute to light strange baryon
	production at the inclusive level, and is partially compensated
	by the further addition of thermodynamical fragmentation, which
	simultaneously raises $\Xi/\Lz$.
	
	A measurement with ALICE\,3 in Run~5 should be able to distinguish
	models with and without color-string junctions in the charm
	sector with reasonable luminosity.
	The pattern of results seen here -- a clear $z$-dependence in the
	charm sector, simple level shifts in the strange sector, and
	angular features that are visible only at the first branching --
	gives several complementary handles for testing how hadronization
	works inside jets.
	    Comparing these observables to predictions from other hadronization models, such as cluster fragmentation, alternative junction prescriptions, or dynamic charm-hadronization models, would be a natural next step that would test whether the junction-based picture identified here still holds up against these competing explanations.
	
	\ack
	The Author would like to thank Antonio Ortiz for the insightful comments provided during the preparation of the manuscript. 
	This work has been supported by the Hungarian National Research,
	Development and Innovation Office (NKFIH) under contract numbers
	NKKP ADVANCED 25-153456 and 2025-1.1.5-NEMZ\_KI-2025-00002, the
	Wigner Scientific Computing Laboratory (WSCLAB), and the HUN-REN
	Cloud.
	

\end{document}